\documentclass[notitlepage,12pt]{report}

\newcommand{\be}{\begin{equation}}
\newcommand{\ee}{\end{equation}}
 
\newcommand{\bea}{\begin{eqnarray}}
\newcommand{\eea}{\end{eqnarray}}

\usepackage{csquotes}

\begin{document}

\centerline{\bf 
A Commentary on {\it The Knowledge Machine} by 
Michael Strevens}

\centerline{July 24, 2021} 

\hfil

{\footnotesize 

\centerline{Manuel~Ortega-Rodr\1guez}
 
\centerline{Escuela de F\1sica, 
Universidad de Costa Rica} 

\centerline{11501-2060 San Jos\'e, Costa Rica;} 

\centerline{manuel.ortega@ucr.ac.cr} }

\begin{abstract}
\noindent
We offer a few comments derived from a careful reading of  
Michael Strevens' book {\it The Knowledge Machine} (TKM), 
with an emphasis on extensions for future work.
We believe this book goes well beyond traditional accounts 
of scientific change, and offers thus many 
insights into new research.
\end{abstract}



\hfil  

\noindent{\bf TKM's Argument: The Iron Rule} 

\hfil 

Disclaimer. This brief take is certainly no substitute for a careful reading of
the book's arguments, and it is cast in my own words 
and reflects my own understanding (that of a practicing
theoretical physicist).

After thousands of years of flirting with technology and ideas, 
humankind invented a type of activity that was qualitatively different
from anything that came before.
This activity, which we now call ``modern science,'' 
may be thought of as an emergent property, 
in the sense of complex systems, 
and it is something that one could even dare to say that 
just came about, rather than being gradually and purposedly crafted
over the centuries.
After that, as is the case with complex systems, there has been an invisible hand
which steadily moves science towards its objectives.

This invention is a game whose rules guarantee that discussions
are self-perpetuated in a sustainable and fertile fashion, in a way that is just not possible in
pre-scientific, philosophical discussions.

In this game, 
you are trying to win an argument, with the rule that ideology not be brought into the discussion.
Ideology (i.e. your beliefs) in this game is effectively treated as a taboo.

To be sure, 
you can have your beliefs---and as a scientist you pretty much have to,
in order to engine your scientific {\it private} reasoning---but 
your beliefs are never to be offered as argumentation.
The only possible argumentation rests on explanatory power that is directly 
related to a form of agreed-upon testing.

A universal agreement (and this is part of the emergent character of science) 
is thus reached
in which it is undisputed what kind of testing counts as corroboration or refutation 
and, equally important, in which the
loser part always has the possibility of a ``rematch'' 
by challenging one or more auxiliary conditions, a situation again to be decided 
by testing.  

These particular workings of the process are collectively 
called the Iron Rule (IR) and guarantee
sustainability and movement towards science's objectives. 
The IR is self-sustained in the sense that there is no controlling entity 
overseeing the process.

The result of all this is noise in the short time scale and science in the long time scale, 
i.e. an activity that looks grainy and idiosyncratic locally but that overall converges to something 
that we could call the truth, or at least some kind of truth
(something {\it useful}).

The author also makes the case that the only reason science appeared in 17th century Europe and 
not, say, in Ancient Greece or China is because of the appearance of a complex, multi-layered social 
system in Europe that worked as a metaphor for the structure of science as an activity. 
(To be specific, in the form of a {\it totemic} relation between the pairs 
religion : nationality, on one hand, 
and private scientific reasoning : public argumentation, on the other.) 




\hfil

\noindent{\bf Comments, in no particular order}

\begin{enumerate}
         
         
         \item I believe there is a lot to be learned  
         by casting TKM's argument in a {\bf complex systems} framework.
         Think of science as a tiny corner of the ``phase space'' 
         (the space of possibilities) of human ideas and actions. 
         By chance alone, 
         science would have taken maybe a million years to appear spontaneously. 
         But change the ``external'' conditions, 
         and you have an emergent property, a {\bf phase transition} into science.
         (In this case, the external conditions are the multi-layeredness of the social 
         structure in 17th century Europe.) 
         
         {\it Opportunity for new research:} Is the scenario described in TKM consistent with what we know about complex systems and phase 
         transitions?\footnote{beyond the usual approach of network models 
         and simulations} 
         Just as in a spontaneously crystallized undercooled liquid, once the new phase 
         (science) appears, it is somewhat 
         immune to reversing back to the previous phase 
         (philosophical thought). 
  
         \item One of the book's main conclusions, the one that recommends that science be let alone, 
         not trying to fix it because it is just fine, 
         could easily be read by some as a conservative stance, supporting the status quo 
         of white men in science and going against the advancement of women in science, for example. 
          
         {\it Opportunity for new research:}  Far from it, I think the book's ideas can be taken 
         as an invitation to rethink how to best catalyze the entrance of women, and other groups, 
         into science, not by forcing things but rather by using analogies from other complex systems, 
         acknowledging that simplistic and naive solutions probably will not work. 
         
         \item In addition to seeing the argument of TKM under the light of complex systems, 
         one could also bring in what is known about {\bf traditional knowledge} in general.
         In this case, however, traditional knowledge lies not in the contents
         {\it but in the method itself}. It emerges as the system (humanity's ideas) finds  
         a loophole (a new structure) in order to organize itself differently.  
         
         {\it Opportunity for new research:} Can the workings of the IR be understood 
         in terms of what we know in general about traditional knowledge systems around the world
         (and through time)?
         
         
         \item The IR may explain, in very simple terms, why the ideas related
         to landscape and multiverse physics are so problematic within the physics community.
         At times, critics of these ideas seem to know intuitively that something is being
         violated, that there is something truly unscientific there, 
         but failing when trying to articulate it convincingly.
     
         {\it Opportunity for new research:} Can the IR explain in simple terms why
         multiverse physics is outside science, for example by arguing that there is
         too little in explanatory power and too much in metaphysical argumentation? 
         This could also be used to {\it guide} future work in the field.
         
         \item Is the IR still applicable to the science of complex 
         systems,\footnote{Note that in this item complex systems is a possible field of study within 
         science. This is different from what is discussed in item 1, where {\it science itself} is
         considered a complex system, akin to a living organism.}
         as in the 
         study of the collective behavior of molecules or animals? 
         As complexity differs from traditional physics, for example in precluding 
         predictions, one may ask whether the IR applies to it.
         
         {\it Opportunity for new research:} Are there any limitations on the IR 
         when dealing with complexity?  Does the IR lose effectiveness in the context
         of a more nuanced explanatory power, such as the one found in complexity thinking?  


         \item A more nuanced stance for the IR can be sought in which the ``all explanation, 
         no ideology'' stance 
         would be replaced with something more subtle, along the lines of Lewis' 
         idea of ``first and second things.'' 
         

         According to this author, if one puts second things first and first things second, then you lose both,
         a typical simple example being running and health; once running becomes more important than health (as in an obsession with running), you
         can easily become unhealthy and stop running.

         {\it Opportunity for new research:} Can a more nuanced stance be taken, 
         in which ideology is not eliminated but merely made sure it is always second? 
         That would give some leeway to metaphysical commitments (such as beauty)  
         in the same sense that a pinball player can tilt slightly the machine as long as she doesn't go too far. 
         (Multiverse physics presumably went too far.)  But what does it mean exactly to be second in this
         context? How do you measure it?



         \item TKR's argument of the multi-layeredness of society (nationality, religion), 
         a feature unique to Western societies in the 17th century, as a working metaphor
         for the construction of the structure of science as an activity, could be 
         critically contrasted with Joseph Henrich's ideas on the particularities of 
         European kinship relations for the same purpose.
         
         {\it Opportunity for new research:} Does Henrich's ideas merit a revision of TKR's 
         argument?  A possibility is that both causes worked in tandem, reinforcing each 
         other.\footnote{Henrich's ideas are important for those who (like Steven Shapin) 
         think that {\it trust} 
         was a key element in the development of modern science.} 
         

         
         \item An intriguing idea is whether the arguments expounded in TKR can be used 
         to cast predictions or not, predictions about how science works in our present
         society. It is not even clear that the IR can be put to use in such a way.
         
         {\it Opportunity for new research:} One could try, for example, to formulate
         some sort of algorithm that would determine which candidates for scientific activity
         will prosper or stagnate. This algorithm would say something like ``multiverse
         physics'' is not scientific for such and such reasons. 
         
         
         \item Could it be possible that the maxim of leaving  
         science alone worked well in the past, but as the social world is changing
         we need to update that view?  Analogies can be found in the realms of
         economics and law, where the rapid emergence of the internet has resulted 
         in a situation of destructive catch-up. (Think of how pre-internet legal systems
         were not really equipped to handle the subtleties of privacy issues 
         in internet platforms.) 
         
         {\it Opportunity for new research:} Does science need some minor help 
         to keep on being effective?  If so, what kind of help?  Can analogies from 
         other complex systems be helpful?
         
         \item TFK argues that a new form of activity, modern science, appeared
         in the 17th century with no precedent and with features that make it 
         qualitatively different from its predecessors. 
         Should we expect then the appearance of 
         a new form of science, let us call it Science Plus, or Super-Science,  
         which is to science what science is to pre-science?
         
         {\it Opportunity for new research:} What would this extrapolated Science Plus look like?
         Could it be ``artificially'' 
         induced,\footnote{as with a seed crystal in the case of undercooled liquids} 
         or would we need some sort of (possibly uncontrollable) societal change for that?   
         Is modern physics already 
         this Science Plus? That is, do we have already an Iron Rule 2.0 in  
         Peter Galison's ``intercalated periodization'' 
         explanation of progress in 
         physics (with theoretical and experimental physics as separate, symbiotic disciplines)? 
         
  
\end{enumerate}

\newpage  

\noindent{\bf Acknowledgements}

\hfil  

I would like to thank Michael Strevens for his insightful feedback, and in particular
for the comment in footnote 3.
This work was supported by
grant 805-A4-125 of the Universidad de Costa Rica’s Vicerrector\1a de 
Investigación and the CIGEFI. 

\hfil 

\hfil  
 
\noindent{\bf References}

\hfil 


\noindent 
Michael Strevens, {\it The Knowledge Machine;
How Irrationality Created Modern Science} (New York: Liveright Publishing, 2020).  

\hfil 

\noindent 
Clive S. Lewis, ``First and Second Things,'' in {\it God in the Dock} 
(Grand Rapids, MI: William B. Eerdmans Publishing, 1970), pp. 278-280.

\hfil 

\noindent 
Joseph Henrich,
{\it The WEIRDest People in the World; How the West Became Psychologically Peculiar and Particularly Prosperous} 
(New York: Farrar, Straus \& Giroux, 2020).

\hfil 

\noindent 
Peter Galison, ``Trading Zone; Coordinating Action and Belief,'' in
{\it The Science Studies Reader}, M. Biagoli, ed. 
(New York: Routledge, 1999). See the discussion around diagram 10.3
on p.$\, 143$.




\end{document}